# Dynamics and recovery of genuine multipartite Einstein-Podolsky-Rosen steering and genuine multipartite nonlocality for a dissipative Dirac system via Unruh effect

Wen-Yang Sun[1], Dong Wang[1, 2] and Liu Ye[1,*]

[1] *School of Physics & Material Science, Anhui University, Hefei 230601, People's Republic of China*

[2] *CAS Key Laboratory of Quantum Information, University of Science and Technology of China, Hefei 230026, People's Republic of China*

**Abstract:** In this paper, we investigate the dynamics behaviors of genuine multipartite Einstein-Podolsky-Rosen steering (GMS) and genuine multipartite nonlocality (GMN), and explore how to recover the lost GMS and GMN under a mixed decoherence system. Explicitly, the decoherence system can be modeled by that a tripartite Werner-type state suffers from the non-Markovian regimes and one subsystem of the tripartite is under a non-inertial frame. The conditions for steerable and nonlocal states can be obtained with respect to the tripartite Werner-type state established initially. GMS and GMN are very fragile and vulnerable under the influence of the collective decoherence. GMS and GMN will vanish with growing intensity of the Unruh effect and the non-Markovian reservoir. Besides, all achievable GMN's states are steerable, while not every steerable state (GMS's state) can achieve nonlocality. It means that the steering-nonlocality hierarchy is still tenable and GMN's states are a strict subset of the GMS's states in such a scenario. Subsequently, we put forward an available methodology to recover the damaged GMS and GMN. It turns out that the lost GMS and GMN can be effectively restored, and the ability of GMS and GMN to suppress the collective decoherence can be enhanced.

## Introduction

Bell nonlocality [1] and Einstein-Podolsky-Rosen (EPR) steering [2] are two fundamental

---

[*] Corresponding author: yeliu@ahu.edu.cn

features of quantum mechanics. Actually, both of them are two different types of concepts in the region of quantum theory, even if they are indeed intimately related. Bell nonlocality is a property of quantum correlation that goes beyond the paradigm of local realism [3-5]. On the basis of the famed Bell theorem [4], correlations between the results of measurements of local observables on some quantum states might be nonlocal. This nonlocal character of quantum states can be revealed by violating the corresponding Bell-type inequalities [6-12]. On the other hand, the phenomenon of EPR steering was introduced by Schrödinger [13] to analyze the EPR-paradox in 1935. Conceptually, EPR steering describes the ability that an observer can instantaneously affect a remote system by making use of local measurements. Later on, Wiseman, Jones and Doherty have formulated EPR steering within an operational way in conformity during a quantum information task [14]. EPR steering can be understood as an intermediate form of quantum correlation between entanglement [15] and Bell nonlocality [16, 17] within modern quantum information theory, and is very useful to explore the relation between these concepts.

Additionally, two basic questions concerning EPR steering are its detection and quantification. EPR steering can be detected by the violation of EPR steering inequalities [18-24], the violation of which provides a sufficient condition for a given quantum state to be steerable. Derived for both continuous and discrete variables systems, such some EPR steering inequalities can be obtained by utilizing entropy uncertainty relation [21, 22]. The first EPR steering criterion based on the Heisenberg uncertainty relation for conditional measurements of the amplitude $x$ and phase $p$ quadrature of light fields was derived by Reid in 1989 [25], which can be applicable to continuous variable systems, as considered in EPR's original argument. Furthermore, the quantification of EPR steering has attracted more and more attention in the past few years [19, 26-29], which results in several useful EPR steering measures, such as EPR steerable weight [19, 29] and EPR steering robustness [26]. What's more, growing attention has been directed to EPR steering owing to its potential applications within quantum information processing [30-40], for example, one-sided device-independent quantum key distribution [32, 34, 36, 37], quantum secure communication [30, 38], directional quantum teleportation [39, 40], and entanglement assisted sub-channel discrimination [26].

To date, bipartite EPR steering has been theoretically studied in different perspectives [41-43] and a variety of quantum systems [44-46], and has been experimentally testified for both continuous [47] and discrete variables [48]. In addition, He and Reid [49] presented a theoretical proof to probe genuine multipartite EPR steering (GMS) in 2013, which is a natural multipartite extension of the original EPR paradox. They also derived an extraordinary form for multipartite EPR steering, called collective multipartite EPR steering, where the predetermination for the quadrature of a single mode could be enough to violate the inferred uncertainty relation only when all the remaining parties cooperatively work. Later, Wang *et al.* [50] have formalized the collective tripartite EPR steering in terms of local hidden state model and provided EPR steering inequalities that act as signatures and suggest how to optimize collective tripartite EPR steering in specific optics-based scheme.

In a realistic regime, quantum systems unavoidably suffer from decoherence caused by the interaction between the systems and its surrounding environments. Typically, environments usually are classified into two categories, viz., Markovian and non-Markovian ones [51-56]. Specifically, Markovian environment is characterized by leading to the degradation of the correlation of a quantum object [55]. By contrast, as the system interacts with a normative non-Markovian environment [56, 57], the revival of quantum entanglement after a finite time period of the entire absence can be realized [58]. On the other hand, understanding quantum phenomena in a relativistic frame is of basic importance because the realistic quantum systems are essentially non-inertial. As a consequence, considering between external environments and non-inertial effect are indispensable and significant under a realistic regime in the course of quantum information processing. However, in the past years, there have been a few authors to examine the steerability of bipartite [8, 59-61] and tripartite [62-64] states in the local environments only under the cases of inertial systems. And some authors [8, 65] have investigated the relations among entanglement, EPR steering and Bell nonlocality for bipartite states under the inertial frames. Therefore, in this paper, we will concentrate on investigating GMS and genuine multipartite nonlocality (GMN) introduced by Svetlichny [12] within non-Markovian regimes and the influence of the Unruh effect [66, 67] triggered by particle's acceleration when the particle is under a non-inertial frame. Herein, we mainly investigate the dynamics behaviors of GMS and GMN and the relationship between them under the collective influence of non-Markovian environment and the Unruh effect, and then focus on how to effectively recover the lost quantum correlations (GMS and GMN).

The remainder of this paper is organized as follows. In Sec. II, we introduce the criteria to identify the GMS and the GMN in multipartite quantum systems, respectively. Then, we investigate the dynamics behaviors of GMS and GMN for the initial tripartite Werner-type states under the mixed decoherence caused by the collective influence of the non-Markovian environments and the Unruh effect in Sec. III. In Sec. IV, we put forward a scheme to restore the lost GMS and GMN in the mixed decoherence systems. Finally, we end up our paper with a brief conclusion.

## II. Quantum measures in multipartite systems: genuine multipartite EPR steering and genuine multipartite nonlocality

*Genuine multipartite EPR steering (GMS)*

The genuine multipartite EPR steering was introduced by He and Reid [49] in 2013. According to their viewpoint, similar with the Svetlichny model, to prove EPR steering, one needs to falsify a description of the statistics based on a model where the averages can be written as

$$\langle X_1 X_2 X_3 \rangle = P_1 \sum_R p_R^{(1)} \langle X_2 X_3 \rangle_R \langle X_1 \rangle_{R,\rho} + P_2 \sum_R p_R^{(2)} \langle X_1 X_3 \rangle_R \langle X_2 \rangle_{R,\rho} + P_3 \sum_R p_R^{(3)} \langle X_1 X_2 \rangle_R \langle X_3 \rangle_{R,\rho}, \quad (1)$$

where $\sum_R p_R^{(i)} = 1$ and the subscript $\rho$ denotes that the averages are consistent with those of a quantum density matrix. No such constraint is made for the moments $\langle X_j X_k \rangle_R$, written without the subscript $\rho$, and $\langle X \rangle = \text{Tr}(\rho X)$. This model is one in which the system is in a probabilistic mixture, with probabilities $P_1, P_2, P_3$ of the three bipartitions of the system.

If we consider that the tripartite system is a three-qubit system, with the usual Pauli spin operators defined for each site. The uncertainty relation for spin implies $(\Delta \sigma_x^{(k)})^2 + (\Delta \sigma_y^{(k)})^2 \geq 1$, $(\Delta \sigma_z^{(k)})^2 + (\Delta \sigma_y^{(k)})^2 \geq 1$ and $(\Delta \sigma_z^{(k)})^2 + (\Delta \sigma_x^{(k)})^2 \geq 1$ for each site $k = 1, 2, 3$. We can see that the approach given in Ref. [49] will be applied, to give conditions for GMS. One can define

$$\begin{aligned} S_{\text{I}} &:= \langle [\Delta(\sigma_z^{(1)} - \sigma_z^{(2)})]^2 \rangle + \langle [\Delta(\sigma_x^{(1)} + \sigma_y^{(2)} \sigma_y^{(3)})]^2 \rangle \geq 1, \\ S_{\text{II}} &:= \langle [\Delta(\sigma_z^{(2)} - \sigma_z^{(3)})]^2 \rangle + \langle [\Delta(\sigma_x^{(2)} + \sigma_y^{(1)} \sigma_y^{(3)})]^2 \rangle \geq 1, \\ S_{\text{III}} &:= \langle [\Delta(\sigma_z^{(3)} - \sigma_z^{(1)})]^2 \rangle + \langle [\Delta(\sigma_x^{(3)} + \sigma_y^{(2)} \sigma_y^{(1)})]^2 \rangle \geq 1. \end{aligned} \quad (2)$$

Assume that the systems are described by the bipartitions $\{A_s, B_s\}$. For a three-qubit state, the possible bipartitions $\{A_s, B_s\}$ are $\{23,1\}_s$, $\{13,2\}_s$ and $\{12,3\}_s$. Actually, the inequality $S_\text{I}$ is implied by bipartition $\{23,1\}_s$, the inequality $S_\text{II}$ is implied by $\{13,2\}_s$, and the inequality $S_\text{III}$ is implied by $\{12,3\}_s$. Hence, one can obtain the GMS-inequality of the tripartite qubit-state

$$GMS_I(\rho) := \{S_\text{I} + S_\text{II} + S_\text{III} \geq 1\}. \tag{3}$$

Violation of the GMS-inequality expressed in Eq. (3) is sufficient to show GMS. In addition, the value of GMS-inequality is smaller, which means that the ability of GMS is stronger.

### *Genuine multipartite nonlocality (GMN)*

GMN can be tested employing a method pioneered by Svetlichny [12]. Accordingly, the Svetlichny inequality will be used to test the GMN [68]. The Svetlichny inequality for a tripartite qubit-state $\rho$ can be expressed as

$$\mathbb{S} := |\text{Tr}\,(S \cdot \rho)| \leq 4, \tag{4}$$

with

$$S = ABC + ABC' + AB'C + A'BC - A'B'C - A'BC' - AB'C' - A'B'C', \tag{5}$$

where the measurement operators $I$ and $I'$ correspond to the measurements on the qubit $I$ ($I = A, B, C$) with the primed and unprimed terms denoting two different measurements directions. The measurements operators on the second qubit differ by $\vartheta_I$ from those performed on the first one

$$\begin{pmatrix} I \\ I' \end{pmatrix} = \begin{pmatrix} \cos\vartheta_I & -\sin\vartheta_I \\ \sin\vartheta_I & \cos\vartheta_I \end{pmatrix} \begin{pmatrix} A \\ A' \end{pmatrix}, \tag{6}$$

Herein, we define $A = \sigma_y$ and $A' = \sigma_x$, the corresponding measurements operators for parts $A$, $B$, and $C$ can be given by

$$\begin{aligned}
&A = \sigma_y \otimes \mathbb{I}_2 \otimes \mathbb{I}_2,\ A' = \sigma_x \otimes \mathbb{I}_2 \otimes \mathbb{I}_2, \\
&B = \mathbb{I}_2 \otimes (\cos\vartheta_B\ \sigma_y - \sin\vartheta_B\ \sigma_x) \otimes \mathbb{I}_2, \\
&B' = \mathbb{I}_2 \otimes (\cos\vartheta_B\ \sigma_x + \sin\vartheta_B\ \sigma_y) \otimes \mathbb{I}_2, \\
&C = \mathbb{I}_2 \otimes \mathbb{I}_2 \otimes (\cos\vartheta_C\ \sigma_y - \sin\vartheta_C\ \sigma_x), \\
&C' = \mathbb{I}_2 \otimes \mathbb{I}_2 \otimes (\cos\vartheta_C\ \sigma_x + \sin\vartheta_C\ \sigma_y),
\end{aligned} \tag{7}$$

where $\sigma_x$ and $\sigma_y$ are Pauli matrices, $\mathbb{I}_2$ is a 2×2 identity matrix, $\vartheta_B$ and $\vartheta_C$ are rotation angles. For conveniently, we define that the GMN-inequality for the tripartite qubit-state can be expressed as

$$GMN_I(\rho) := \left\{ \frac{1}{4} | \text{Tr } (S \cdot \rho) | \leq 1 \right\}, \quad (8)$$

If a tripartite qubit-state violates the GMN-inequality, which means that the state satisfies GMN.

## III. Dynamics of GMS and GMN for the tripartite Werner-type states within the mixed decoherence systems

Suppose that there are three observers, say, Alice, Bob and Charlie, sharing an initial three-qubit state (tripartite Werner-type states) in the form [69-71]

$$\rho_{Werner-type} = (1-V)(|GHZ\rangle\langle GHZ|) + \frac{V}{8} I_8, \quad 0 \leq V \leq 1, \quad (9)$$

where $|GHZ\rangle = \frac{\sqrt{2}}{2}(|000\rangle + |111\rangle)$, and $I_8$ is an 8×8 identity matrix. Based on Eqs. (3) and (8), we can obtain its GMS-inequality $\frac{3}{16} + \frac{9}{4}V \geq 1$ and GMN-inequality $\sqrt{2}(1-V) \leq 1$, respectively. As shown in Fig. 1, one can see that the tripartite Werner-type state is a steerable state in the case of $0 \leq V < \frac{13}{36}$, and the tripartite Werner-type state satisfies GMN for $0 \leq V < \frac{2-\sqrt{2}}{2}$. Besides, the maximally entangled state ($V=0$) is the maximally GMN and GMS states. Thus, we can conclude that for the whole set of three-qubit states it holds that $GMN \Rightarrow GMS$ implying a hierarchy according to which all GMN's states are steerable, while no every GMS's state can satisfy GMN.

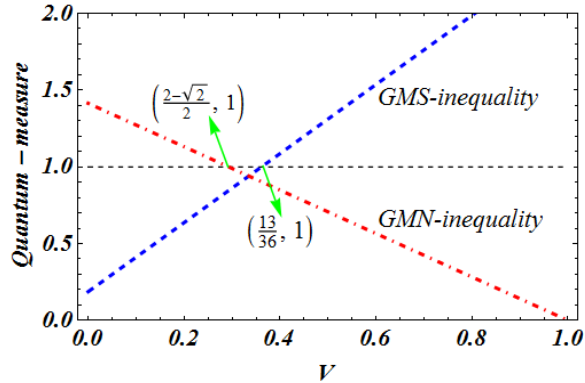

**Fig. 1** (Color online) Two quantum-measure (GMS and GMN inequalities) as function of the state parameters $V$ when they initially share a tripartite Werner-type state.

Then, considering the three-qubit systems expressed in Eq. (9), each locally and independently interacts with a zero-temperature reservoir. Herein, the single "qubit+reservoir" Hamiltonian [72, 73] can be given by

$$H = \omega_0 \sigma_+ \sigma_- + \sum_i \omega_i b_i^\dagger b_i + \sum_i \left( \sigma_+ g_i b_i + \sigma_- g_i^* b_i^\dagger \right), \quad (10)$$

where $g_i$ are the coupling constants, $\omega_0$ denotes the transition frequency of the two-level qubit and $\sigma_\pm$ are the raising and lowering operators. Besides, $b_i^\dagger (b_i)$ is the modes' creation (annihilation) operator and the index $i$ labels the field modes of the reservoir with frequencies $\omega_i$. Via Ref. [74, 75], the dissipative reservoir of non-Markovian regime at zero temperature can be modeled as a multi-mode electromagnetic field in a lossy cavity with the effective spectral density $J(\omega) = \gamma_0 \lambda^2 / \{2\pi[\lambda^2 + (\omega_0 - \omega)^2]\}$, with $\gamma_0$ and $\lambda$ define the decay rate of the excited state of the qubit and the spectral width of the reservoir, respectively. Specifically, their relative magnitudes typically determine a non-Markovian regime ($\lambda < 2\gamma_0$) and a Markovian ($\lambda > 2\gamma_0$) regime, respectively [72]. We can discover that the action of reservoir over single qubit $S$ can be represented by the following quantum maps

$$\begin{aligned}
|0\rangle_S |0\rangle_E &\rightarrow |0\rangle_S |0\rangle_E, \\
|1\rangle_S |0\rangle_E &\rightarrow \sqrt{P_t} |1\rangle_S |0\rangle_E + \sqrt{1-P_t} |0\rangle_S |1\rangle_E.
\end{aligned} \quad (11)$$

And the corresponding Kraus operators are given by

$$K_0 = \begin{pmatrix} 1 & 0 \\ 0 & \sqrt{P_t} \end{pmatrix}, \quad K_1 = \begin{pmatrix} 0 & \sqrt{1-P_t} \\ 0 & 0 \end{pmatrix}. \quad (12)$$

Within this regime, the function $P_t$ can be given by [72]

$$P_t = e^{-\lambda t} \left[ \cos\left(\frac{\delta t}{2}\right) + \frac{\lambda}{\delta} \sin\left(\frac{\delta t}{2}\right) \right]^2, \quad (13)$$

where $\delta = \sqrt{2\gamma_0 \lambda - \lambda^2}$. The time-evolution of the initial tripartite Werner-type states under the non-Markovian regime can be expressed by the trace-preserving quantum operation $\xi(\rho)$, which is $\xi(\rho) = \sum_i K_i \rho K_i^\dagger$ with the Kraus operators satisfying trace-preserving condition $\sum_i K_i K_i^\dagger = I$.

After they undergone the non-Markovian regimes. Alice and Bob are as inertial observers who remain stationary at a flat space-time region, while Charlie is a non-inertial observer traveling with a uniform acceleration $a$ in the Rindler space-time. Under the non-inertial frame, due to the constant acceleration, Charlie travels on a hyperbola constrained in the region I which is causally disconnected from the other region II under the Rindler space-time. Besides, since the field modes restricted in different regions cannot be connected, the information loss for the accelerated observer results from a thermal bath. From the perspective of accelerated observers, the vacuum and excited states can be written as [76-84]

$$|0\rangle_U \to \cos\chi |0\rangle_I |0\rangle_{II} + \sin\chi |1\rangle_I |1\rangle_{II},$$
$$|1\rangle_U \to |1\rangle_I |0\rangle_{II},$$
(14)

respectively. Here, the dimensionless parameter $\chi$ is defined by $\cos\chi = (e^{-\frac{2\pi}{a/\omega}} + 1)^{-\frac{1}{2}}$, $a$ is Charlie's acceleration, and $\omega$ is the frequency of the Dirac particle [85-87]. It should be noted that in [81] the so-called single-mode approximation (SMA) has been imposed—which, however, is not correct for general states. And beyond the single-mode approximation [81, 88, 89] can be simplified to the SMA under the case of specific. Additionally, the corresponding Kraus operators for Unruh noise can be expressed as

$$K_0 = \begin{pmatrix} \cos\chi & 0 \\ 0 & 1 \end{pmatrix}, \quad K_1 = \begin{pmatrix} 0 & 0 \\ \sin\chi & 0 \end{pmatrix}.$$
(15)

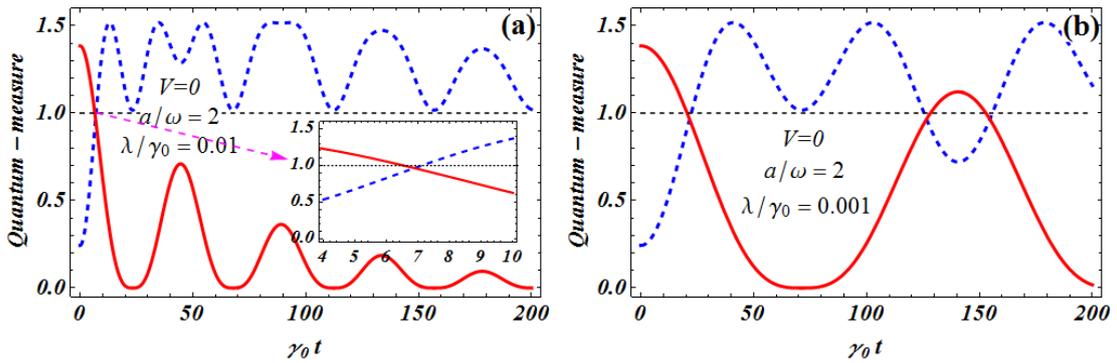

**Fig. 2** (Color online) A variety of quantum-measure (GMS-inequality (blue dashed line), and GMN-inequality (red solid line)) as function of $\gamma_0 t$ in terms of different value $\lambda/\gamma_0$ for $V=0$, $a/\omega=2$. (a) $\lambda/\gamma_0 = 0.01$, (b) $\lambda/\gamma_0 = 0.001$.

Thereby, via above-mentioned these processes, we can obtain that the non-zero elements of the final evolution state $\rho(t)_{Werner-type}$ are

$$\rho_{11}(t) = \frac{1}{8}(P_t - 2)\left[P_t(4 + (-4 + 3V)P_t) - 4\right]\cos^2\chi,$$

$$\rho_{22}(t) = \frac{1}{8}\left[P_t(4 + P_t(4V - 8 + 4P_t - 3VP_t)) + (P_t - 2)\left[P_t(4 + (3V - 4)P_t) - 4\right]\sin^2\chi\right],$$

$$\rho_{33}(t) = \frac{1}{8}P_t\left[4 + P_t(4V - 8 + 4P_t - 3VP_t)\right]\cos^2\chi,$$

$$\rho_{44}(t) = \frac{1}{16}P_t\left(4 + (3V - 4)P_t^2 + \left[P_t(8 - 4P_t + V(3P_t - 4)) - 4\right]\cos(2\chi)\right),$$

$$\rho_{55}(t) = \frac{1}{8}P_t\left(4 + P_t(4V - 8 + 4P_t - 3VP_t)\right)\cos^2\chi,$$

$$\rho_{66}(t) = \frac{1}{16}P_t\left(4 + (3V - 4)P_t^2 + \left[P_t(8 - 4P_t + V(3P_t - 4)) - 4\right]\cos(2\chi)\right),$$

$$\rho_{77}(t) = \frac{1}{8}P_t^2\left(4 - 4P_t + V(3P_t - 2)\right)\cos^2\chi,$$

$$\rho_{88}(t) = \frac{1}{8}P_t^2\left[(4 - 3V)P_t + (4 - 4P_t + V(3P_t - 2))\sin^2\chi\right],$$

$$\rho_{18}(t) = \rho_{81}(t) = \frac{1}{2}(1 - V)P_t^{3/2}\cos\chi.$$

(16)

Employing Eq. (8), the expression of GMN-inequality can be given by

$$\sqrt{2}(1-V)P_t^{3/2}\cos\chi \leq 1. \quad (17)$$

If Eq. (17) is violated, we can obtain that the system reveals GMN under the influence of mixed decoherence (non-Markovian regimes and the Unruh effect). Besides, by using Eq. (3), we can obtain an analytical expression of GMS-inequality for the states $\rho(t)_{Werner-type}$.

In terms of the above discussion, we can plot the GMS-inequality and GMN-inequality of $\rho(t)_{Werner-type}$ as a function of $\gamma_0 t$ in terms of different value $\lambda/\gamma_0$ for $V = 0$, $a/\omega = 2$ in Fig. 2. From these figures, some results can be obtained as follows: (1) When $\lambda/\gamma_0 = 0.01$, the GMS and GMN will vanish with the increasing $\gamma_0 t$, but the revivals of GMS and GMN cannot achieve after periods of $\gamma_0 t$ in the presence of Unruh noise in the Fig. 2 (a). However, when $\lambda/\gamma_0 = 0.001$, the GMS and GMN will achieve again after periods of $\gamma_0 t$ in the Fig. 2 (b). It means that if the GMS and GMN could achieve again with the increasing $\gamma_0 t$, $\gamma_0$ should be much larger than $\lambda$ (i.e., $\lambda/\gamma_0$ should be small enough). (2) One can obtain that all achievable GMN's states are steerable, while no every GMS's state can achieve GMN. Namely, we correctly verify the steering-nonlocality hierarchy in the case of the tripartite Werner-type states under the influence of mixed decoherence.

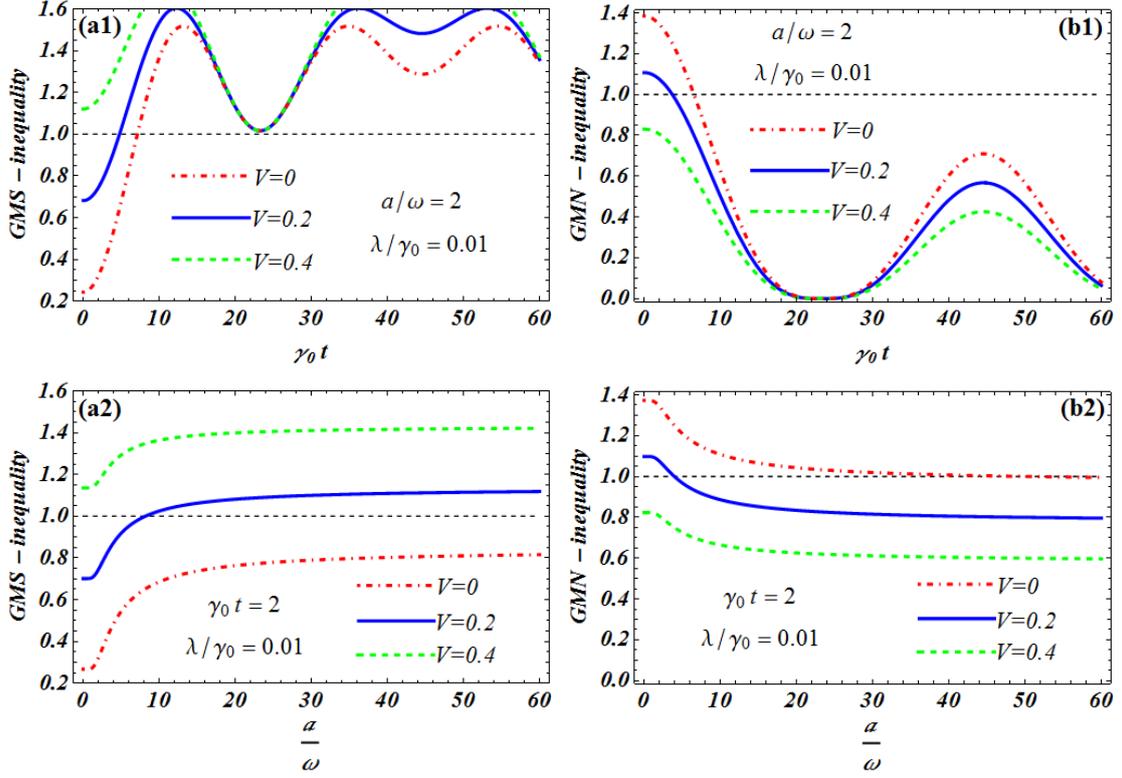

**Fig. 3** (Color online) **(a1)**, **(b1)** GMS-inequality and GMN-inequality as function of $\gamma_0 t$ in terms of different state parameter $V$ for $\lambda/\gamma_0 = 0.01$, $a/\omega = 2$, respectively. **(a2)**, **(b2)** GMS-inequality and GMN-inequality as function of $a/\omega$ in terms of different state parameter $V$ for $\lambda/\gamma_0 = 0.01$, $\gamma_0 t = 2$, respectively.

Next, in order to better understand the relations among the GMS (GMN), the state parameter $V$, the parameters $a/\omega$ and $\gamma_0 t$ under the mixed decoherence systems for $\lambda/\gamma_0 = 0.01$, we plot the graphs of relations among them in Fig. 3. As shown in Fig. 3, we can obtain that the GMS-inequality increases with the increase of the state parameter $V$ and the parameter $a/\omega$, and increases at first and then decreases (subsequently, the cycle of ups and downs) with the increase of the parameter $\gamma_0 t$. However, the dynamics behaviors of GMN-inequality are the diametrically opposite. Through these phenomena, we can obtain a conclusion that GMS and GMN will disappear with the increase of the parameters $V, a/\omega$ and $\gamma_0 t$, but the GMS and GMN cannot realize again with the increase of $\gamma_0 t$ in the presence of Unruh noise. All evolutionary states could be employed to realize GMS with the increasing $a/\omega$ for $\gamma_0 t = 2$ in Fig.3 (a2), when the state parameter is equal to zero. And the maximally steerable state is the maximally nonlocal state. Additionally, the steering-nonlocality hierarchy still maintains in the case of the tripartite Werner-type states under the collective influence of the non-Markovian environments and the Unruh effect.

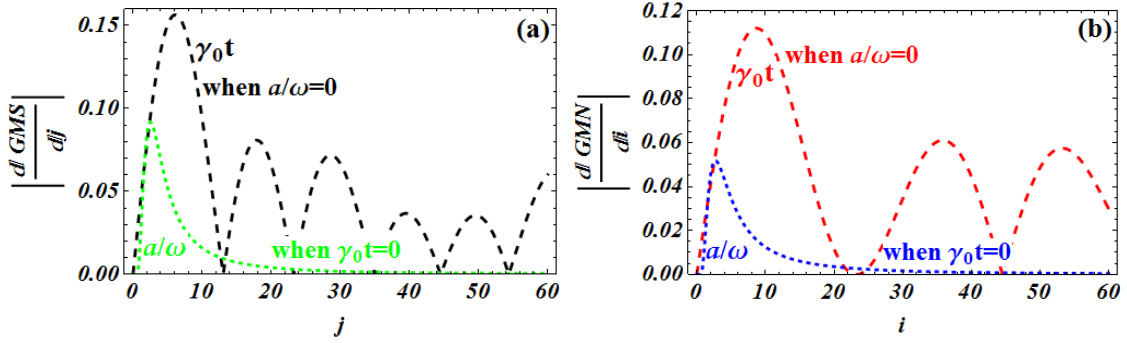

**Fig. 4** (Color online) **(a)** The absolute value of first-order partial derivative of GMS-inequality with the different parameters $j, i$ $(i(j) = \gamma_0 t, a/\omega)$, when $a/\omega=0$ or $\gamma_0 t=0$. **(b)** The absolute value of first-order partial derivative of GMN-inequality with the different parameters $j, i$ $(i(j) = \gamma_0 t, a/\omega)$, when $a/\omega=0$ or $\gamma_0 t=0$.

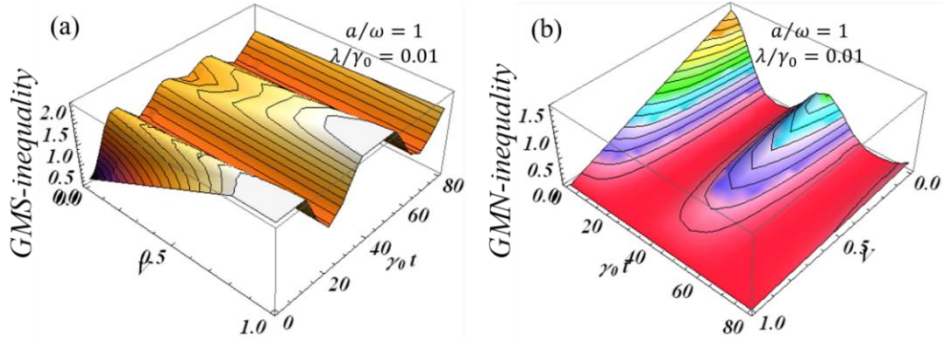

**Fig. 5** (Color online) **(a)** The GMS-inequality as functions of the state parameter $V$ and $\gamma_0 t$ for $\lambda/\gamma_0 = 0.01$, $a/\omega = 1$. **(b)** The GMN-inequality as functions of $\gamma_0 t$ and state parameter $V$ for $\lambda/\gamma_0 = 0.01$, $a/\omega = 1$.

Furthermore, we find that GMS and GMN are very fragile and vulnerable under the influence of the mixed decoherence. The Unruh effect and non-Markovian environments can seriously influence the GMS and GMN. However, the impact of non-Markovian environments on the GMS and GMN is stronger than that of the Unruh effect. To clarify the result, we illustrate mathematically the relationship between absolute value of first-order partial derivative of GMS-inequality (or GMN-inequality) and different parameters ($\gamma_0 t, a/\omega$) in Fig. 4. As shown in Fig. 4 (a), the absolute value of first-order partial derivative of GMS-inequality-$\gamma_0 t$ ($a/\omega=0$) is greater than that of GMS-inequality-$a/\omega$ when $\gamma_0 t=0$. Fig. 4 (b) shows that the impact of the Unruh effect on GMS and GMN is weaker than that of non-Markovian environments. If the state parameter is equal to zero, we can realize the maximal GMS and GMN, and also find that GMS and GMN is very sensitive under the influence of non-Markovian environments in Fig. 5.

## IV. Recovering Genuine Multipartite Einstein-Podolsky-Rosen Steering and Genuine Multipartite Nonlocality

As is well known, quantum correlations will be decreased or even disappeared when the qubits undergo the dissipative channels. Thus, it is important to protect quantum correlations in open quantum systems. Actually, the vulnerability of a qubit suffered from the environment is caused by its spontaneous emission, whose rate is proportional to the weight of its excited-state ($|1\rangle$) component [90, 91]. Because the weak measurement (WM) operations can reduce the excited-state component weight [91], via utilizing prior WM on three qubits synchronously, which changes the weight between the excited-state and the ground-state, and partially collapses the states from $|111\rangle$ to $|000\rangle$ in our scheme and make the qubits more robust against decoherence. Hence, we can use WM to better suppress the effect of the mixed decoherence (non-Markovian environments and the Unruh effect) on GMS and GMN. The corresponding WM can be written as a non-unitary quantum operation

$$M_{wk}(m_A, m_B, m_C) = \begin{pmatrix} 1 & 0 \\ 0 & \sqrt{1-m_A} \end{pmatrix} \otimes \begin{pmatrix} 1 & 0 \\ 0 & \sqrt{1-m_B} \end{pmatrix} \otimes \begin{pmatrix} 1 & 0 \\ 0 & \sqrt{1-m_C} \end{pmatrix}, \quad (18)$$

where $m_A, m_B$ and $m_C$ are WM strengths. After three qubits interact with the non-Markovian environments, we perform weak measurement reversal (WMR) operations on Alice and Bob, while Charlie travels with a uniform acceleration *a* in a non-inertial frame. On the other hand, the WMR as another non-unitary operation can be given by

$$M_{rev}(m_{r_A}, m_{r_B}, I_c) = \begin{pmatrix} \sqrt{1-m_{r_A}} & 0 \\ 0 & 1 \end{pmatrix} \otimes \begin{pmatrix} \sqrt{1-m_{r_B}} & 0 \\ 0 & 1 \end{pmatrix} \otimes \begin{pmatrix} 1 & 0 \\ 0 & 1 \end{pmatrix}. \quad (19)$$

where $m_{r_A}$ and $m_{r_B}$ are WMR strengths. For simplicity, we here assume $m_A = m_B = m_C = m$ and $m_{r_A} = m_{r_B} = m_r$. For clarity, the physical scheme sketch of the total system is depicted in Fig. 6, where the dissipative environment is turned to the non-Markovian regimes. Thus, the non-zero elements of the final evolution states $\rho(t)^w_{Werner-type}$ are

$$\rho^w_{11}(t) = \frac{\left(4-3V+3V(m-1)(P_t-1)+3V(1-m)^2(P_t-1)^2-(3V-4)(m-1)^3(P_t-1)^3\right)\cos^2\chi(1-m_r)^2}{\Theta},$$

$$\rho^w_{22}(t) = \frac{(m_r-1)^2(1-m)\left(V+2V(m-1)(P_t-1)-(3V-4)(m-1)^2(P_t-1)^2\right)P_t}{\Theta}$$
$$+\frac{(m_r-1)^2\left(4-3V+3V(m-1)(P_t-1)+3V(m-1)^2(P_t-1)^2-(3V-4)(m-1)^3(P_t-1)^3\right)\sin^2\chi}{\Theta},$$

$$\rho^w_{33}(t) = \frac{(1-m)(1-m_r)\left(V+2V(m-1)(P_t-1)-(3V-4)(m-1)^2(P_t-1)^2\right)p\cos^2\chi}{\Theta},$$

$$\rho^w_{44}(t) = \frac{(1-m_r)P_t(1-m)^2 P_t\left(4(m-1)(P_t-1)+V(3P_t-2-3m(P_t-1))\right)}{\Theta}$$
$$+\frac{(1-m_r)P_t(1-m)\left(V+2V(m-1)(P_t-1)-(3V-4)(1-m)^2(1-P_t)^2\right)\sin^2\chi}{\Theta},$$

(20-a)

$$\rho^w_{55}(t) = \frac{(1-m)(1-m_r)\left(V+2V(m-1)(P_t-1)-(3V-4)(1-m)^2(1-P_t)^2\right)P_t\cos^2\chi}{\Theta},$$

$$\rho^w_{66}(t) = \frac{(1-m_r)P_t(1-m)^2 P_t\left(4(m-1)(P_t-1)+V(3P_t-2-3m(P_t-1))\right)}{\Theta}$$
$$+\frac{(1-m_r)P_t(1-m)\left(V+2V(m-1)(P_t-1)-(3V-4)(1-m)^2(1-P_t)^2\right)\sin^2\chi}{\Theta},$$

(20-b)

$$\rho^w_{77}(t) = \frac{(1-m)^2 P_t^2\left(4(1-m)(1-P_t)+V(3m(1-P_t)+3P_t-2)\right)\cos^2\chi}{\Theta},$$

$$\rho^w_{88}(t) = \frac{(1-m)^2 P_t^2\left((3V-4)(1-m)P_t\cos^2\chi+(2V-4+4m-3Vm)\sin^2\chi\right)}{\Theta},$$

$$\rho^w_{18}(t) = \rho^w_{81}(t) = \frac{4(V-1)(1-m)^{3/2}(m_r-1)P_t^{3/2}\cos\chi}{\Theta},$$

with

$$\Theta = (m-2)\left(m(4+(3V-4)m)-4\right)(m_r-1)^2+(m-1)^2\left(4-4m+V(3m-2)\right)m_r^2 P_t^2$$
$$-2(m-1)\left(m(8-4m+V(3m-4))-4\right)(m_r-1)m_r P_t.$$

(21)

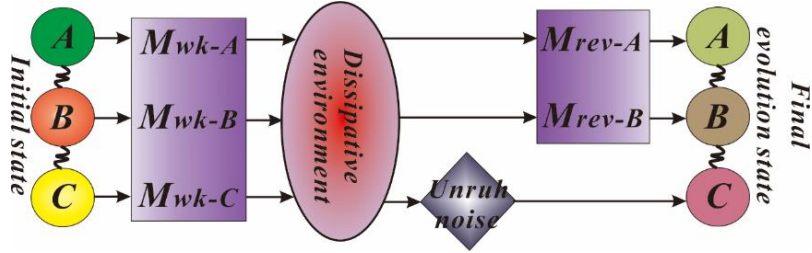

**Fig. 6** (Color online) Schematic diagram of systems: recovering the GMS and GMN of the tripartite Werner-type state is under the non-Markovian regimes and the Unruh noise.

Now, we have at hand two control parameters $m$ and $m_r$, which we are able to tune to manipulate the qubits' correlations for a certain purpose at any time during the evolution. If we would like to obtain the best effect for GMS and GMN recovery, we need an optimal WMR strength. According to Refs. [91-94], the optimal WMR strength $m_r^o$ can be given by

$$m_r^o = 1 - \frac{(1-m)^2(4-4m+V(3m-2))P_t^2}{\sqrt{(1-m)^2(4-4m+V(3m-2))P_t^2\left((m-2)(m(4+(3V-4)m)-4)-\Xi\right)}}, \quad (22)$$

where $\Xi = 2(m-1)(m(8-4m+V(3m-4))-4)P_t + (1-m)^2(4-4m+V(3m-2))P_t^2$. We can find that the optimal WMR strength has nothing to do with the Unruh noise. Then, by choosing the optimal WMR strength, and substituting Eq. (20) into Eqs. (3) and (8), respectively, we can obtain the analytical expressions of GMS-inequality and GMN-inequality for the final state $\rho(t)_{Werner-type}^w$, respectively.

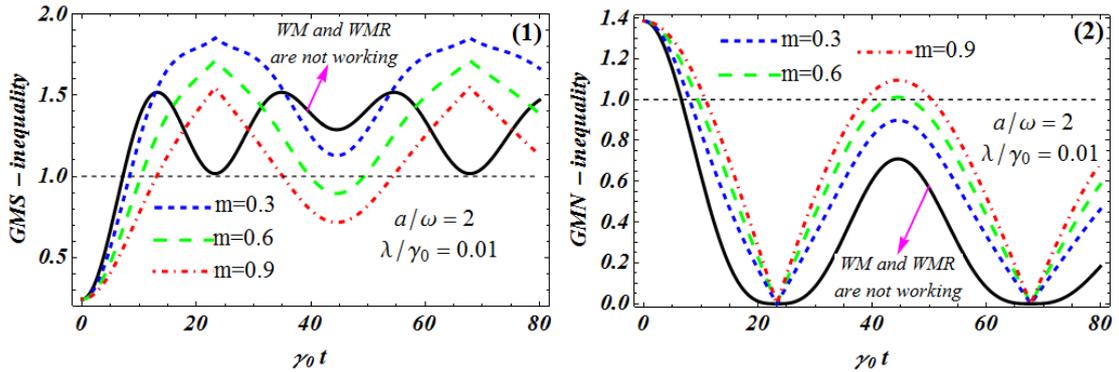

**Fig. 7** (Color online) **(1)**, **(2)** GMS-inequality and GMN-inequality as function of $\gamma_0 t$ in terms of different WM strength $m$ for $\lambda/\gamma_0 = 0.01$, $a/\omega = 2$, $V = 0$, respectively.

As shown in Fig. 7, if we do not implement the WM and WMR operations, the capacity of GMS and GMN suppressing to the mixed decoherence are very weak. And GMS and GMN cannot achieve again with the increasing $\gamma_0 t$ in the presence of Unruh noise. However, if we implement the WM and WMR operations and choose the optimal WMR strength for Eq. (22), it is found that the ability of GMS and GMN to resist the mixed decoherence become stronger with the increasing the WM strength, and GMS and GMN can achieve again with the increase of $\gamma_0 t$. The results indicate that WM and WMR operations can effectively recover the damaged GMS and GMN caused by the collective influence of the non-Markovian environments and the Unruh effect, and the larger the WM strength, the better the GMS and GMN recovery.

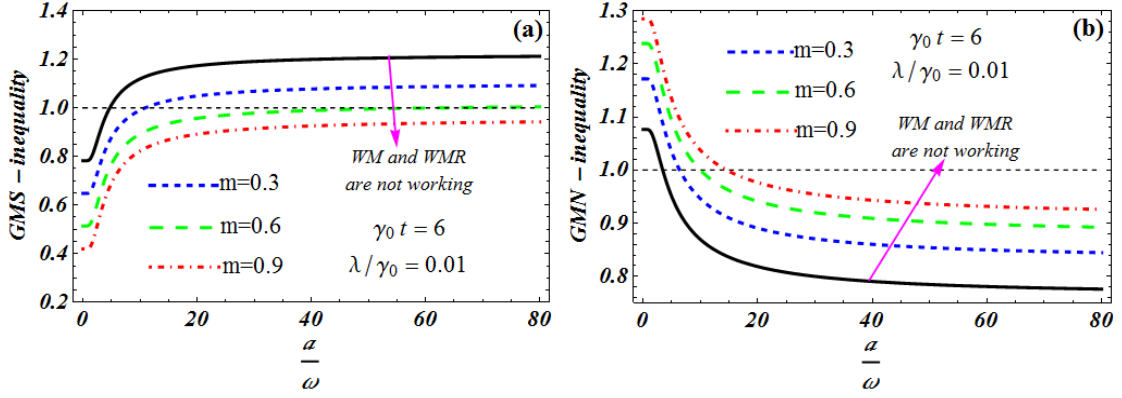

**Fig. 8** (Color online) **(a)**, **(b)** GMS-inequality and GMN-inequality as function of $a/\omega$ in terms of different WM strength $m$ for $\lambda/\gamma_0 = 0.01$, $\gamma_0 t = 6$, $V = 0$, respectively.

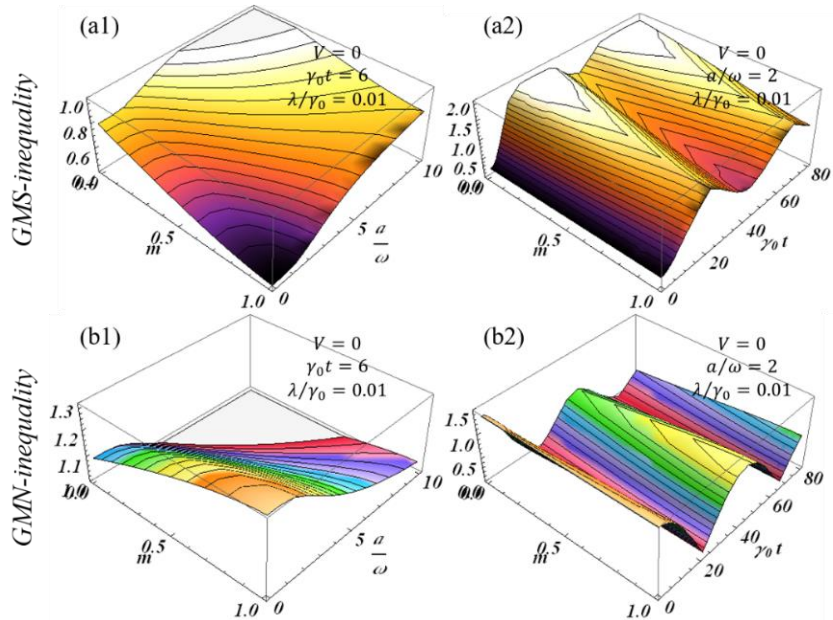

**Fig. 9** (Color online) **(a1), (b1)** The GMS-inequality and GMN-inequality as functions of the WM strength $m$ and $a/\omega$ for $\lambda/\gamma_0 = 0.01$, $V = 0$, $\gamma_0 t = 6$, respectively. **(a2), (b2)** The GMS-inequality and GMN-inequality as functions of the WM strength $m$ and $\gamma_0 t$ for $\lambda/\gamma_0 = 0.01$, $V = 0$, $a/\omega = 2$, respectively.

Subsequently, we display how GMS and GMN vary with $a/\omega$ and WM strength $m$ with $V = 0$, $\gamma_0 t = 6$ and $\lambda/\gamma_0 = 0.01$ in Fig. 8. From these figures, we can conclude that both GMS and GMN of the given evolution state can be remarkably recovered by implementing WMs operation. Besides, when the strength of WM is big enough, the given evolution state could be utilized to achieve GMS, whatever the value of $a/\omega$ is. However, GMN is dramatically destroyed by the Unruh effect caused by the acceleration of Charlie, and suffers from a "sudden death" with the growing intensity of the Unruh effect in the the non-Markovian regimes in Fig. 8 (b), even if the strength of WM is big enough, which means that for the same operating strength of WMs, restoring GMS is easier than restoring GMN. These results also verify that the steering-nonlocality hierarchy is reasonable in our scenario. The above phenomenon is also illustrated in Fig. 9. These results entirely illustrated that our scheme can effectively restore the lost GMS and GMN by utilizing WM and WMR operations. And it is disclosed that the larger the strength of WM is, the better effectiveness the scheme has.

## V. Conclusions

To conclude, we have investigated the collective influence of the non-Markovian environments and the Unruh effect on GMS and GMN, and explore how to recover the lost GMS and GMN for a tripartite Werner-type state. For the initial tripartite Werner-type state, we can obtain that the tripartite Werner-type state is a steerable state in the case of $0 \leq V < \frac{13}{36}$. Besides, the tripartite Werner-type state can satisfy GMN for $0 \leq V < \frac{2-\sqrt{2}}{2}$. The Unruh effect and non-Markovian environments can seriously influence the GMS and GMN. However, the impact of the Unruh effect on GMS and GMN is weaker than that of non-Markovian environments. The results show that GMS and GMN decrease with the growing intensity of the Unruh effect and the non-Markovian thermal bath, and GMS and GMN are very fragile and vulnerable under the collective influence of the Unruh effect and the non-Markovian regimes. It means that the ability of GMS and GMN suppressing to the mixed decoherence are very weak. Additionally, if the GMS and GMN could achieve again with the increasing $\gamma_0 t$ in the presence of the Unruh noise, only when $\gamma_0$ is much larger than $\lambda$. Interestingly, it has been found that the steering-nonlocality hierarchy is still tenable in our scenario. Subsequently, in order to restore the damaged GMS and GMN, we put forward an available methodology to recover the damaged GMS and GMN by making use of quantum partially collapsing measurements (the pre-measurement (WM) and the post-measurement (WMR) operations). It turns out that the damaged GMS and GMN can be effectively restored. And the ability of GMS and GMN to suppress the mixed decoherence can be enhanced. Moreover, it is verified that the larger the WM strength is, the better the effectiveness of our scheme is. Consequently, we believe that our work might be helpful to understand the dynamics behaviors of GMS and GMN, and also be effective to recover the lost GMS and GMN within an open system.

## Acknowledgments

This work was supported by the National Science Foundation of China under Grant Nos. 11575001 and 61601002, Anhui Provincial Natural Science Foundation (Grant No. 1508085QF139) and Natural Science Foundation of Education Department of Anhui Province (Grant No. KJ2016SD49), and also the fund from CAS Key Laboratory of Quantum Information (Grant No. KQI201701).